# Vulnerability of multilayer network systems to system-wide lesions

Olexandr Polishchuk, Dmytro Polishchuk
Laboratory of Modeling and Optimization of Complex Systems,
Pidstryhach Institute for Applied Problems of Mechanics and Mathematics, National Academy of Sciences of Ukraine
Lviv, Ukraine
od_polishchuk@ukr.net

**Abstract.** On the basis of structural and flow models of multilayer network system (MLNS), the main structural and functional importance indicators of separate layers in the process of intersystem interactions are calculated. With the help of influence and betweenness parameters of separate layers, their role as generators, final receivers and transitors of intersystem flows is determined. Using the introduced importance indicators, effective scenarios of system-wide targeted attacks on MLNS are developed. The problem of the scale of targeted attacks and non-target lesions consequences for the network system (NS) and determination of its protection and sensitivity to various negative influences is analyzed. The advantages of flow-based approach over the structural ones during the study of vulnerability of NS and MLNS to heterogeneous internal and external negative influences are shown.

## I. Introduction

The main types of negative internal and external influences on real-world complex network systems and processes of intersystem interactions were analyzed in the article [1]. Among such influences, targeted attacks that have artificial nature and intruder who carries out this attack, and non-target lesions of complex systems of natural or artificial origin, which do not have such intruder, were primarily distinguished. Such negative influences can have a local, group or system-wide nature and be aimed at damaging both the structure and operation process of network and multilayer network systems. History knows many examples when local lesions of real-world NS and MLNS grew into group ones, and group lesions into system-wide ones (transition of epidemics into pandemics [2], local military conflicts into world wars [3], development of marginal extremist views into national ideologies [4], spread of agricultural pests and invasion processes [5, 6], etc.). In the theory of complex networks (TCN), strategies for protection against system-wide attacks and non-target lesions of NS and MLNS were paid relatively little attention until recently, and only Covid-19 and the impact of the russian-ukrainian war on the world community have shown that such damages are still quite real and humanity generally unprepared to counteract their spread and overcome the consequences [7, 8]. The scale of targeted attack or system non-target lesion is usually determined by the quantity of directly damaged (destroyed) elements that should be removed from its structure. This can significantly reduce the importance of problem, because as a result of Covid-19 and russian aggression, not only those who fell seriously ill or were injured, died or perished, but the whole society was suffered. Taking into account the quantity of consequentially injured by negative influence NS and MLNS elements can significantly increase the damaged area. That is, the problem of more accurate classification and understanding both the lesion itself and its consequences arises. The concept of system protection is closely related to the scale of its lesions. During the russian-ukrainian war, there was an increment of quantity of downed missiles and drones from 10% at the beginning of the aggression to 85% or more in its second year [9]. Increasing the level of system protection can significantly contribute to reducing the quantity of both directly damaged and consequentially injured MLNS elements. The purpose of article is to develop effective scenarios of system-wide targeted attacks on MLNS, methods for evaluation the scale of damage and system protection.

## II. SYSTEM-WIDE LESIONS AND MODELS OF MULTILAYER NETWORK SYSTEMS

System-wide negative influences will be considered such actions in relation to the system, during which all its elements are damaged to one degree or another. It is clear that the larger the attack target, the more difficult it is to achieve the expected result. Therefore, system-wide lesion is often attempted through successive group attacks, such as missile strikes on oil depots in Ukraine in the spring and summer of 2022, which led to fuel shortages in the country, or attacks on energy infrastructure facilities during 2022-24, which resulted in introduction of electricity blackout schedules for almost all of its consumers. Scenarios of system-wide attacks can be based on provoking so-called cascade phenomena, which most often occur in MLNS with hierarchical network structure [10]. Examples of such lesions are the blackouts caused by cascading phenomena in the power grids of Italy and USA in 2003 [11]. The road transport system of a large city can collapse as a result of accident at one of the intersections [12], etc. And although these are examples of non-target lesions caused by equipment malfunctions or traffic accidents, they can become the basis for the organization of system-wide targeted attacks. Therefore, special attention is now being paid to the problem of deployment of cascade phenomena in network and multilayer network systems [13]. The blocking of sea and air transport due to the danger of destruction of ships and planes can be considered as general system attack on the transport system of Ukraine. Permanent air strikes cause considerable damage to the moral and psychological state of the population of Ukraine, and in the regions of continuous missile or artillery attacks, local residents are in panic state. Non-target system-wide lesions include the Covid-19 pandemic, global warming, the decrease in biodiversity of the Earth's flora and fauna, etc. Targeted attacks and non-targeted lesions of MLNS can affect both a separate system layer and several layers or a multilayer system at a whole. The monograph [14] showed how, based on the structural and flow models of multilayer network system, it is possible to determine the importance indicators of its elements and subsystems, to use these indicators for construction the effective scenarios of targeted attacks on the structure and operation process of multilayer system, and to evaluate the consequences of such attacks. We will apply these models to determine the importance indicators of separate MLNS layers and build effective scenarios of targeted attacks on the process of intersystem interactions at a whole.

### A. A structural model of intersystem interactions

The structural model of intersystem interactions is described by multilayer networks (MLNs) and represented in the form [15]

$$G^M = \left( \bigcup_{m=1}^{M} G_m, \bigcup_{m,k=1,\ m\neq k}^{M} E_{mk} \right),$$

where $G_m = (V_m, E_m)$ determines the structure of $m^{th}$ network layer of MLN; $V_m$ and $E_m$ are the sets of nodes and edges of network $G_m$ respectively; $E_{mk}$ is the set of connections between the nodes of $V_m$ and $V_k$, $m \neq k$, $m,k = \overline{1,M}$, and $M$ is the quantity of MLN layers. The set $V^M = \bigcup_{m=1}^{M} V_m$ will be called the total set of MLN nodes, $N^M$ – the quantity of $V^M$ elements. In this paper, we consider partially overlapped MLN [16], in which connections are possible only between nodes with the same numbers from the total set of nodes $V^M$ (Fig.1). This means that each node can be an element of several systems and perform one function in them, but in different ways. Then the set $E^M = \bigcup_{m=1}^{M} E_m$ will be called the total set of MLN edges, $L^M$ – the quantity of elements of $E^M$.

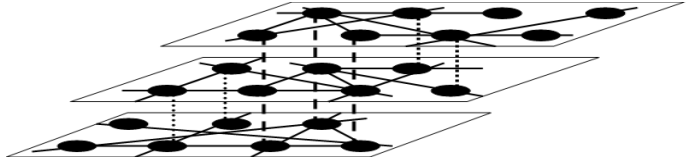

Fig. 1. Example of structure of partially overlapped multilayer network

Nodes through which interlayer interactions are carried out will be called MLNS transition points. The quantity of connections $K^M$, which are carried out through the transition points from layer to layer, in this case is calculated by the formula

$$K^M = \sum_{i=1}^{N^M}(k_i - 1)$$

in which the value $k_i$ is equal to the quantity of layers which includes a node $n_i$, $i = \overline{1, N^M}$, of the total set of nodes $V_m$, $m = \overline{1, M}$.

Multilayer network $G^M$ is fully described by an adjacency matrix

$$\mathbf{A}^M = \{\mathbf{A}^{km}\}_{m,k=1}^M, \qquad (1)$$

in which the blocks $\mathbf{A}^{mm}$ determine the structure of intralayer and blocks $\mathbf{A}^{km}$, $m \neq k$, – interlayer interactions. Values $a_{ij}^{km} = 1$ if the edge connected the nodes $n_i^k$ and $n_j^m$ exists, and $a_{ij}^{km} = 0$, $i, j = \overline{1, N^M}$, $m, k = \overline{1, M}$, if such edge don't exists. Blocks $\mathbf{A}^{km} = \{a_{ij}^{km}\}_{i,j=1}^{N^M}$, $m, k = \overline{1, M}$, of matrix $\mathbf{A}^M$ are determined for the total set of MLN nodes, i.e. the problem of coordination of node numbers is removed in case of their independent numbering for each layer.

### B. A flow model of intersystem interactions

In monograph [14], a method of decomposition of multidimensional (multiflow) MLNS [17] into monoflow multilayer systems was proposed and a flow model of such systems was developed, which makes it possible to calculate the main local and global functional characteristics of the elements of such formations. To determine the indicators of functional importance of monoflow MLNS's components, we will use the flow model [1], which is described by the adjacency matrix $\mathbf{V}^M(t)$, the elements of which are determined by the volumes of flows that have passed through the edges of MLN (1) for the period $[t-T, t]$ up to the current moment of time $t \geq T > 0$:

$$\mathbf{V}^M(t) = \{V_{ij}^{km}(t)\}_{i,j=1, \; k,m=1}^{N^M \quad M}, \qquad (2)$$

$$V_{ij}^{km}(t) = \tilde{V}_{ij}^{km}(t) \Big/ \max_{s,g=\overline{1,M}} \max_{l,p=\overline{1,N^M}} \{\tilde{V}_{lp}^{sg}(t)\},$$

where $\tilde{V}_{ij}^{km}(t)$ is the volume of flows that passed through the edge $(n_i^k, n_j^m)$ of multilayer network for the time period $[t-T, t]$, $i, j = \overline{1, N^M}$, $k, m = \overline{1, M}$, $t \geq T > 0$. It is obvious that structure of matrix $\mathbf{V}^M(t)$ completely coincides with the structure of matrix $\mathbf{A}^M$. The elements of MLNS flow adjacency matrix (2) are determined on the basis of empirical data about movement of flows through MLNS edges. Currently, with the help of modern means of information extraction, such data can be easily obtained for many real-world natural and the vast majority of man-made systems [18]. The matrix $\mathbf{V}^M(t)$ similarly to $\mathbf{A}^M$ also has a block structure, in which the diagonal blocks $\mathbf{V}^{mm}(t)$ describe the volumes of intralayer flows in the $m^{th}$ layer, and the off-diagonal blocks $\mathbf{V}^{km}(t)$, $m \neq k$, describe the volumes of flows between the $m^{th}$ and $k^{th}$ layers of MLNS, $m, k = \overline{1, M}$, $t \geq T > 0$.

Typical scenarios of successive and simultaneous targeted attacks on layers-systems of MLNS are built according to the same principles as corresponding scenarios of such attacks on elements and subsystems of multilayer systems [1] with the difference that the objects of attack are the layers of MLNS. At the same time, the main way to improve the effectiveness of such attacks is the selection of structural and/or functional importance indicators of MLNS layer, the lesion of which will cause the greatest damage to it.

### III. IMPORTANCE INDICATORS OF MLNS SYSTEM-LAYERS

The global structural characteristics of multilayer network, which is generated by monoflow partially overlapped MLNS, can be considered the quantity of layers-networks that are part of it, as well as the dimension, density, diameter, average length of the shortest path, the total quantity of transition points [15], etc.

#### A. Structural importance indicators of system-layers

Let us define some general and most important for the vulnerability analysis of structure of intersystem interactions characteristics of the layers of partially overlapped MLNS, the values of which should be taken into account during construction the scenarios of system-wide targeted attacks:

1) the specific weight $\theta_m$ of the set of nodes of the $m^{th}$ layer in the total set of nodes of partially overlapped MLN, which is determined by the ratio $\theta_m = N_m / N^M$ ;

2) the specific weight $\vartheta_m$ of the set of edges of the $m^{th}$ layer in the total set of edges of partially overlapped MLN, which is determined by the ratio $\vartheta_m = L_m / L^M$ ;

3) input degree of the $m^{th}$ layer, which is equal to quantity of transition point into this layer from all other layers of MLN or total quantity its input interlayer connections;

4) output degree of the $m^{th}$ layer, which is equal to quantity of transition point from this layer into all other layers of MLN or total quantity its output interlayer connections;

5) the specific weight of transition points of the $m^{th}$ layer in the set of all transition points of partially overlapped MLN, which is determined availability of interlayer interactions for this layer, $m = \overline{1, M}$ .

#### B. Flow-based importance indicators of system-layers

The integral indicator of MLNS operation process $s(\mathbf{V}^M(t))$ is equal to the total volumes of flows in monoflow partially overlapped multilayer system during the time period $[t-T, t]$, $t \geq T > 0$, and calculated by the formula

$$s(\mathbf{V}^M(t)) = \sum_{m,k=1}^{M} s(\mathbf{V}^{mk}(t)) ,$$

in which parameters

$$s(\mathbf{V}^{mk}(t)) = \sum_{i,j=1}^{N^M} V_{ij}^{mk}(t)$$

determine the total volumes of intralayer flows in the $m^{th}$ layer, if $k = m$, and volumes of intersystem flows between $m^{th}$ and $k^{th}$ layers, if $k \neq m$, $m, k = \overline{1, M}$ .

Let us determine the most important flow characteristics of system-layers in monoflow partially overlapped multilayer network system, values of which should be taken into account during construction scenarios of system-wide targeted attacks:

1) the specific volumes $V_m^{int}(t)$ of intralayers flows in the $m^{th}$ MLNS layer, which is determined by the ratio

$$V_m^{int}(t) = s(\mathbf{V}^{mm}(t)) / s(\mathbf{V}^M(t)) ;$$

2) the specific volumes $V_m^{ext,out}(t)$ of output flows of the $m^{th}$ layer in MLNS, which reflect its role as a flow generator in multilayer system and are determined by the ratio

$$V_m^{ext,out}(t) = \sum_{k=1, k \neq m}^{M} s(\mathbf{V}^{mk}(t)) / s(\mathbf{V}^M(t)) ;$$

3) the specific volumes $V_m^{ext,in}(t)$ of input flows of the $m^{th}$ layer in MLNS, which reflect its role as a final receiver of flows in multilayer system and are determined by the ratio

$$V_m^{ext,in}(t) = \sum_{k=1, k \neq m}^{M} s(\mathbf{V}^{km}(t)) / s(\mathbf{V}^M(t));$$

4) the specific volumes $\Phi^m(t)$ of transit flows that pass through the $m^{th}$ MLNS layer and will be calculated in the next section, $m = \overline{1,M}$, $t \geq T > 0$.

It should be noted that the layers with the highest values of structural and functional importance indicators listed above can become the primary targets of successive system-wide attacks on MLNS and be used during construction the scenarios for such attacks.

### IV. INFLUENCE AND BETWEENNESS PARAMETERS OF MLNS SYSTEM-LAYERS

Such global flow characteristics of nodes and subsystems of MLNS were determined in the monograph [14], as parameters of their output and input influence and betweenness. These parameters make it possible to calculate the importance of separate components of multilayer system as generators, final receivers and transitors of flows and allow us to build effective scenarios of successive and simultaneous group targeted attacks on the process of intersystem interactions. However, for the formation of effective scenarios of successive system-wide attacks, it is advisable to determine the influence and betweenness parameters of separate MLNS system-layers.

#### A. Influence parameters of system-layers

Output strength of influence of the $m^{th}$ layer as generator of flows on the $l^{th}$ MLNS layer as final receiver of these flows during the time period $[t-T,t], t \geq T > 0$, is calculated by the formula

$$\xi^{m,l,out}(t) = \sum_{i=1}^{N^M} \xi_i^{m,l,out}(t) / N^M, \quad \xi^{m,l,out}(t) \in [0,1], \quad (3)$$

in which the parameter $\xi_i^{m,l,out}(t)$ determines the output strength of influence of the $i^{th}$ node of $m^{th}$ layer on the $l^{th}$ MLNS layer, i. e. the total volume of flows, which generates in this node and directs for final receiving in the nodes of $l^{th}$ layer [14]. It is obvious, that the value $\mathbf{V}^{ml}(t)$ is calculated as a sum of elements of $i^{th}$ column of the block $\mathbf{V}^{ml}(t)$ of the flow adjacency matrix (2). Domain $R^{m,l,out}(t)$ of output influence of the $m^{th}$ layer on the $l^{th}$ MLNS layer are determined as union of influence domains of nodes-generators of flows of the $m^{th}$ layer on the nodes – final receivers these flows in $l^{th}$ MLNS layer, and the power of this influence $p^{m,l,out}(t)$ is equal to the ratio of guantity of elements of the domain $R^{m,l,out}(t)$ to value $N^M$. Similarly, the input strength $\xi^{m,l,in}(t)$, domain $R^{m,l,in}(t)$ and power $p^{m,l,in}(t)$ of influence of the $m^{th}$ layer as final receiver of flows on the $l^{th}$ MLNS layer as generator of these flows are determined during the time period $[t-T,t], t \geq T > 0$, $m \neq l$, $m,l = \overline{1,M}$.

The output strength of influence of the $m^{th}$ layer as generator of flows on the MLNS at a whole is calculated by formula

$$\xi^{m,out}(t) = \sum_{l=1, l \neq m}^{M} \xi^{m,l,out}(t) / (M-1), \quad \xi^{m,out}(t) \in [0,1],$$

in which the value $\xi^{m,l,out}(t)$ is calculated according to (3) and domain of this influence is determined by the ratio

$$R^{m,out}(t) = \bigcup_{l=1, l \neq m}^{M} R^{m,l,out}(t).$$

The power of output influence of the $m^{th}$ layer on the MLNS at a whole is equal to the ratio of quantity of element of domain $R^{m,out}(t)$ to the value $N^M$. Similarly, the input strength $\xi^{m,in}(t)$, domain $R^{m,in}(t)$ and power $p^{m,in}(t)$ of influence of MLNS on the $m^{th}$ layer as final receiver of flows are determined during the time period $[t-T,t], t \geq T > 0$. The values of parameters of input and output influence of the $m^{th}$ layer on the $l^{th}$ layer or MLNS at a whole make it possible to quantitatively determine how the lesion of this layer will impact on the operation process of the $l^{th}$ layer and the multilayer network system in general, and how many components of intersystem interactions and to what extent will be affected, $m, l = \overline{1, M}$.

### B. Betweenness parameters of system-layers

The next type of global flow characteristics of each MLNS layer is its betweenness parameters, which determine the importance of layer of the multilayer network system in ensuring the movement of transit flows during intersystem interactions [14]. The measure of betweenness of the $m^{th}$ layer within the entire MLNS during period $[t-T,t], t \geq T > 0$, as transitor of flows is calculated using the formula

$$\Phi^m(t) = \sum_{i=1}^{N^M} \Phi_i^m(t) / N^M, \Phi^m(t) \in [0,1],$$

in which the value $\Phi_i^m(t)$ determines the measure of betweenness of $i^{th}$ transition point of the $m^{th}$ layer within the entire MLNS, i. e. the total volume of flows that pass through the node $n_i^m$ from one layer of multilayer system onto another. Betweenness domain of the $m^{th}$ layer in MLNS at a whole is determined from the ratio

$$M^m(t) = \bigcup_{i=1}^{N^M} M_i^m(t),$$

in which $M_i^m(t)$ denotes the betweenness domain of $i^{th}$ transition point of the $m^{th}$ layer within the entire MLNS, and the betweenness power of the $m^{th}$ layer $N^m(t)$ is equal to the ratio of quantity of elements of domain $M^m(t)$, $[t-T,t], t \geq T > 0$, to the value $N^M$. The values of betweenness parameters of the $m^{th}$ layer in MLNS make it possible to quantitatively determine how the lesion of this layer will impact on the process of intersystem interactions in general and how many, which components of the multilayer system and to what extent will be affected, $m = \overline{1, M}$.

### C. Interaction parameters of system-layers

Based on the parameters of input and output influence, as well as betweenness of the $m^{th}$ layer, we can form a global indicator of interaction of this layer with MLNS in general, namely, parameter $\Xi^m(t)$ of the strength of interaction of the $m^{th}$ layer with multilayer system, which is calculated according to the formula

$$\Xi^m(t) = (\xi^{m,out}(t) + \xi^{m,in}(t) + \Phi^m(t))/3,$$

and defines its overall role in multilayer system as generator, final receiver and transitor of flows; domain $\Omega^m(t)$ of interaction of the $m^{th}$ layer with MLNS is determined by the ratio

$$\Omega^m(t) = R^{m,out}(t) \cup R^{m,in}(t) \cup M^m(t),$$

and the power of interaction of the $m^{th}$ layer with MLNS is equal to the ratio of quantity of elemens of the domain $\Omega^m(t)$, $m = \overline{1, M}$, $t \geq T > 0$, to the value $N^M$. It is clear that interaction parameters of a layer with multilayer network system define the importance of this layer in the process of intersystem interactions and make it possible to quantitatively determine how the lesion of this layer will impact on the MLNS opertion process and how many, exactly which of its components will be influenced and to what extent.

### D. A comprehensive system-wide attack scenario

Let us build a scenario of successive targeted system-wide attack, choosing the parameter of strength of layer interaction with multilayer system as its importance indicator. Such scenario, which achieves the comprehensiveness of attack on the most functionally important MLNS layers, will look like this:
1) compile the list of MLNS layers in order of decreasing values of the parameters of strength of their interaction with multilayer system;

2) remove the first layer from the created list;
3) if the criterion of attack success is reached, then complete the execution of scenario, otherwise go to point 4;
4) since the structure and operation process of multilayer system changes due to removal of certain layer (and its interlayer connections), compile a new list of layers in the order of decreasing values of the parameters of strength of their interaction with MLNS and proceed to point 2.

In this case, it is advisable to choose the reduction of volumes of flows $s(\mathbf{V}^M(t)), t \geq T > 0$ in the multilayer system by a certain predetermined value as the attack success criterion. The next most difficult to ensure protection and overcome the consequences are simultaneous attacks on several or all MLNS layers, which include hybrid wars, sanctions against countries that pose a threat to world security, etc. The scenarios of such attacks are built according to the same principles as those of simultaneous group attacks [1], with the difference that the attack targets are not the most important by certain characteristics the groups of nodes, but the MLNS layers.

## V. TARGETED ATTACKS AND THE SCALE OF SYSTEM LESIONS

The concept of system protection is closely related to determining the scale of targeted attack and its consequences. In particular, it is advisable to distinguish:
1) the scale of planned and executed attack, which is determined by the quantity of targets expected to be hit and the quantity of means used for this purpose, for example, 50 energy infrastructure facilities and 100 cruise and ballistic missiles aimed at them for destruction;
2) the scale of direct lesion, that is, the quantity of targets critically damaged or completely destroyed by the attack, for example, out of 100 launched missiles, 85 were shot down by anti-missile and anti-aircraft defenses, and the remaining 15 missiles hit 10 targets; in this case, the scale of direct damage is equal to the percentage of destroyed objects from the quantity of planned ones;
3) the scale of consequentially injured system elements, i.e. the quantity of objects that suffered certain damage as a result of the attack in addition to those directly damaged, for example, introduction of blackout schedules caused by consecutive and simultaneous attacks on the energy infrastructure of Ukraine.

The scale of direct damage is related to the quantitative indicator of system's protection, which is equal to the percentage of destroyed means of damage from all involved. The scale of mediated lesion, as well as the level of protection against it, is much more difficult to calculate, because it must take into account the disruption of structure and destabilization of the work of all system elements and even the moral and psychological damage caused as a result of the attack. The concept of system sensitivity to the consequences of negative influence can be associated with the scale of indirect lesion, as the ratio of quantity of directly damaged to the quantity of consequentially injured elements. It is obvious that the closer the value of this indicator is to zero, the more sensitive the system is to negative influences, since a small quantity of directly damaged generates a large quantity of consequentially injured NS elements.

The scale of damage can be defined, as in article [1], by comparing the structural and functional models of NS before, during and after a targeted attack. The specific weight of "zeroed" elements and rows and columns of corresponding adjacency matrix determines the quantity of directly damaged edges and nodes of the system. In the structural model, only nodes adjacent to directly damaged nodes and edges connecting them can be considered consequentially injured. This approach sufficiently adequately reflects the level of losses for assortative, for example, biological or social networks [19]. However, for disassortative networks, which include the majority of real-world man-made industrial, economic, financial, transport, information and other systems, it is unsuitable, since many nodes in them are connected not by direct connections, but by paths, determination of which only on the basis of structural model is quite difficult. In this case, the flow model of system is more adequate, namely, the domains and power of input and output influence and betweenness or corresponding interaction domain and power of the set of directly damaged nodes sufficiently unambiguously determine all consequentially injured NS elements.

It is obvious that directly damaged receiver and generator nodes need to be replaced in some way, and for transit nodes, alternative paths of flows movement must be found. All this has a specific financial dimension, which can be used to calculate the level of losses experienced by the system. Indeed, as a result of sanctions against russia due to its aggression in Ukraine, many of the world's leading companies have lost their sales markets (final receivers of flows). Many of them, at least temporarily, had serious problems with the supply of energy resources and raw materials, that is, the problem of replacing generator nodes of certain type of flows arose. The movement of transit flows through the territory of russia was also significantly limited, which required finding alternative and, usually, more expensive and longer ways of movement of these flows [20-22]. Ukraine faced similar circumstances, but as a result of hostilities on its territory (restriction of production, export and import of various products, stoppage of transit flows, etc.). It should also be taken into account that even air alarms, which are announced as a result of the take-off of missile carriers, but are not accompanied by a real strike, also lead to disruptions in the work of educational and medical institutions, as well as many trade, transport, state and industrial enterprises, and generate panic in the population of those regions that constantly suffer from real attacks. Therefore, it makes sense to believe that the nature of attack should be determined not only by the quantity of directly damaged MLNS elements, but also by the scale of indirect losses caused to the system, although it is much more difficult to ensure its protection against such lesion [23-26].

## VI. EVALUATION OF SYSTEM LESION CONSEQUENCES

The picture of targeted attack consequences on the network system, obtained on the basis of structural approach and using its generalized structural degree as importance indicator of a node is shown in fig. 2. The directly damaged NS nodes are marked in black; in dark gray – adjacent to directly damaged (consequentially injured) nodes; in white – undamaged nodes of the network system; the continuous curve delimits the directly damaged NS domain; the dashed curve delimits domain of adjacent consequentially injured network elements.

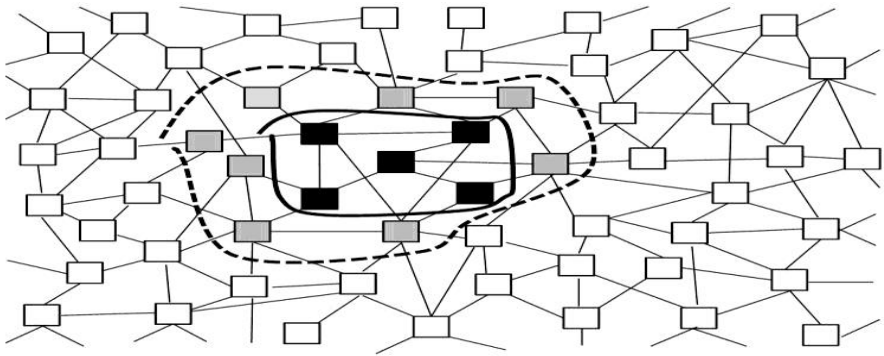

Fig. 2. Lesion consequences obtained on the base of structural approach

The consequences of targeted attack on the network system, obtained on the basis of analysis of its interaction domain (in gray are marked consequentially injured nodes; the dotted curve delimits the consequentially injured NS's domain) is shown in fig. 3. As follows from this picture, the domain of consequentially injured NS elements determined on the basis of flow model can be much larger than domain of adjacent to directly damaged nodes of network system determined on the basis of its structural model [27]. Thus, the comparison of NC flow models before, during, and after negative influence allows us to draw up a sufficiently objective quantitative picture of the lesion level of network system or its separate components as a result of targeted attack or the action of non-target lesion.

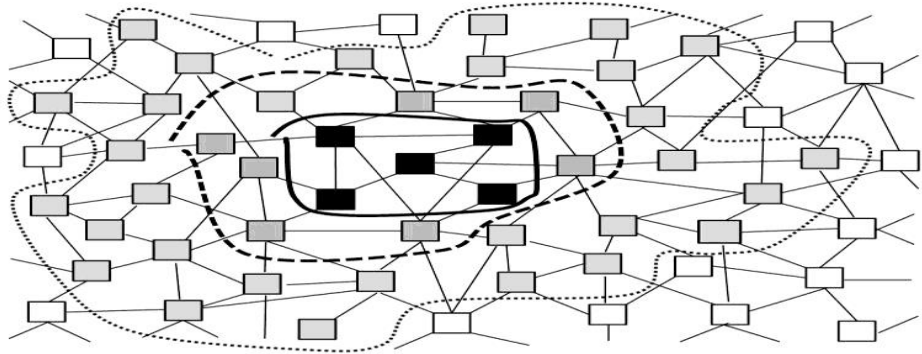

Fig. 3. Lesion consequences obtained on the base of flow interaction parameters analysis

Comparing fig. 2 and 3, it can be reasonably concluded that the flow-based approach allows us to create a much more realistic picture of lesions consequences caused by a certain negative influence than the structural one during which a significant underestimation of the amount of such losses is possible. Similarly, the scale of damage to multilayer networks and systems is determined. Analogous considerations can also be applied to distinguish between direct and indirect effects of non-target system lesions and the process of intersystem interactions in general.

## VII. Conclusions

In 2019-2024, humanity faced two global challenges, the first of which (Covid-19 pandemic) is a vivid example of real-world system-wide non-target lesion, and the second is a targeted attack (attack of the russian federation on Ukraine) and the resulting threat of a global food, energy, and financial crisis and reverse comprehensive sanctions against the aggressor, the negative consequences of which affected almost all countries of the world. Humanity proved to be unprepared for such challenges, but no less dangerous threats remain. Over the past half century, 67% of biological species known to man have disappeared [28], and over the past 20 years, the costs of combating climate disasters have increased 8 times [29]. Currently, scientists know more than 20 viruses of dangerous infectious diseases, the mutations of which can lead to the spread of pandemics, much more catastrophic than Covid-19 [30], the threat of global military and even intercivilizational conflicts is increasing, etc. This confirms the relevance of studying the features of system-wide lesions of complex network and multilayer network systems and developing methods of effective protection against them. Understanding the structural and functional importance of components of the intersystem interactions process makes it possible to choose objects that require priority protection or as soon as possible blocking, as they most contribute to the spread of negative influence. An objective evaluation of the scale of real or potential lesions allows us to develop strategies to protect not only separate MLNS elements, but also the system as a whole, and to prevent the consequences of local targeted attacks and non-target lesions from growing into group and system-wide ones.